\documentclass[draftclsnofoot, onecolumn, 12pt]{IEEEtran}

\usepackage{floatflt}

\usepackage{color}
\usepackage{cite}

\usepackage{graphicx}
\usepackage{amsmath}
\usepackage{amsthm}
\usepackage{amssymb}

\usepackage{verbatim}

\usepackage{psfrag,array}

\usepackage{subfigure}

\usepackage{stfloats}

\usepackage{amsmath}
\usepackage{epstopdf}
\usepackage{algpseudocode}
\usepackage{algorithm}

\newcommand{\p}[1]{\mathop{\mbox{\it p} } }

\renewcommand{\vec}[1]{\ensuremath{\boldsymbol{#1}}}
\newcommand{\be}{\begin{equation}}
\newcommand{\ee}{\end{equation}}
\newcommand{\ba}{\begin{array}}
\newcommand{\ea}{\end{array}}
\newcommand{\bea}{\begin{eqnarray}}
\newcommand{\eea}{\end{eqnarray}}
\newcommand{\bean}{\begin{eqnarray*}}
\newcommand{\eean}{\end{eqnarray*}}
\newcommand{\argmax}{\mathop{\arg\max}}
\newcommand{\argmin}{\mathop{\arg\min}}
\newcommand{\sign}{\mathop{\rm sign}}

\newcommand{\rmh}{^{\rm H}}
\newcommand{\rmt}{^{\rm T}}

\def\NoNumber#1{{\def\alglinenumber##1{}\State #1}\addtocounter{ALG@line}{-2}}

\definecolor{white}{rgb}{1,1,1}

\newtheorem{lemma}{Lemma}

\newtheorem{property}{Property}

\newtheorem{remark}{Remark}

\begin{document}

\title{Modulus Zero-Forcing Detection for MIMO Channels}

\author
{
\begin{tabular}{c}
Sha Hu and Fredrik Rusek \\
 \thanks{The authors are with the Department of Electrical and Information Technology, Lund University, Lund, Sweden (email: \{firstname.lastname\}@eit.lth.se).}
\end{tabular}
}

\maketitle

\begin{abstract}
We propose a modulus arithmetic based zero-forcing (MZF) detector for multi-input multi-output (MIMO) channels. Traditionally, a ZF detector completely eliminates interference from other symbol layers when detecting a particular symbol layer, which results in suboptimal performance due to noise-enhancement. The only constraint for application of our proposed MZF detector is that the transmitter must employ a finite cardinality $M$ quadrature-amplitude-modulation (QAM) alphabet. With that, the modus operandi of the MZF is to allow for integer-valued interference and then remove it by modulus arithmetic operations.

\end{abstract}


\section{Introduction}
We consider a standard multi-input multi-output (MIMO) channel model with a received signal $\tilde{\vec{y}}$ expressed as
\bea \label{complexmodel} \tilde{\vec{y}}=\tilde{\vec{H}}\tilde{\vec{x}} + \tilde{\vec{n}},\eea
where $\tilde{\vec{H}}$, $\tilde{\vec{x}}$ and $\tilde{\vec{n}}$ are the complex-valued MIMO channel, transmitted symbols and Gaussian noise, respectively. 

Given received signal model (\ref{complexmodel}), detecting $\tilde{\vec{x}}$ is referred to as a MIMO detection problem, which has a history that can be traced back about half a century and a review on it can be found in e.g., \cite{YH15}. In general, maximum likelihood (ML) detection \cite{K93} yields optimal performance but with prohibitive complexity when the MIMO dimension and/or the input alphabet has large cardinality. Effective implementations of ML detection, such as sphere-decoding (SD) \cite{GA11, AEZ02} can significantly reduce the complexity, but not overcome an exponential complexity in the number of symbol layers CITE JALDEN'S PAPER. On the other hand, linear detectors \cite{K93} such as zero-forcing (ZF) and linear minimum-mean-square-error (LMMSE), have low complexities, but also suboptimal performances. One direction for improving linear detectors is lattice-aided-reduction (LAR) \cite{WBK04} based approaches, which use lattice-reduction (LR) algorithms, e.g., Lenstra-Lenstra-Lov\'asz (LLL), to find a short and nearly orthogonal basis for the lattice induced by the MIMO channel \cite{LLL}.

Other than the existing approaches \cite{YH15}, as the transmitted symbols are drawn from finite alphabets such as quadrature-amplitude-modulation (QAM) and pulse-amplitude-modulation (PAM) symbols, the modulus can also be used in MIMO detection for improving the detection performance. The modulus operation has been used in Tomlinson-Harashima precoding (THP) \cite{T71, HM72} as suboptimal approximation for dirty-paper coding (DPC) \cite{C83}, and recently it has also be considered in the designs of integer-forcing (IF) receives for MIMO channles~\cite{ZG14, CJ17, OE10}. The IF scheme in \cite{ZG14} requires the transmitter to employ a same lattice code \cite{HC13} for each transmitted layer, which does not apply in most of current communication systems. Besides, when higher-oder modulations such as $M$-PAM are used, designing lattice code over $\mathbb{Z}_M$ is challenging \cite{HC13}. Simpler IF receivers dealing with linear binary codes such as turbo and LDPC are proposed in \cite{OE10, CJ17}, and the designs follow the same as in \cite{ZG14}. One disadvantage of the IF receivers in \cite{ZG14, OE10} is that, each transmit antenna needs a separate encoding/decoding process, which is not the case in practical LTE systems where one codeword is split among transmit-antennas. Moreover, the IF design in \cite{CJ17} needs a separate encoding/decoding process per transmit antenna and per bit-layer in higher-order modulations. Another advantage is that, the receiver has to detect the linear combinations of codewords for all transmit-antennas first, followed by a matrix inversion (over a finite-field) process to recover the original codeword on each layer. 

To overcome these disadvantages in previous IF receiver designs, we consider a new approach to improve linear detection with modulus operation, namely, the proposed modulus ZF (MZF) detection. Note that, MZF is conceptually different from previous IF receivers, although they share quite some similarities. The fundamental difference is that, with MZF, there is no encoding/decoding process needed which are implemented on a finite-filed with IF receivers to recover the transmit symbols (i.e., the linear combinations of codewords across all transmit-antennas). Alternatively, we design MZF detector such that the transmit symbols on each transmit-antenna can be recovered directly by modulating away the interferences from the remaining transmit-antennas, and the symbol detection on different transmit-antenna is independent and fully in parallel. Such a principle simplifies the operation and can be well cooperated into practical systems such as LTE. To achieve such a nice property, with MZF the modulus matrix is carefully designed and optimized according to the specific modulation-order.

\section{Preliminaries}
We start with reviewing the standard ZF detection. Before proceeding, without loss of generality, the matrix $\tilde{\vec{H}}$ is always assumed to be a square matrix, obtained by a QR factorization or padding zero rows to the matrix if necessary. With the following definitions,
\bea \vec{y} = \left[\begin{array}{c} \mathcal{R}\{\tilde{\vec{y}}\} \\ \mathcal{I}\{\tilde{\vec{y}}\}\end{array}\right]\!, \; \vec{x} = \left[\begin{array}{c} \mathcal{R}\{\tilde{\vec{x}}\} \\ \mathcal{I}\{\tilde{\vec{x}}\}\end{array}\right]\!, \; \vec{n} = \left[\begin{array}{c} \mathcal{R}\{\tilde{\vec{n}}\} \\ \mathcal{I}\{\tilde{\vec{n}}\}\end{array}\right]\!, \; \vec{H} = \left[\begin{array}{cc} \mathcal{R}\{\tilde{\vec{H}}\} & -\mathcal{I}\{\tilde{\vec{H}}\}\\ \mathcal{I}\{\tilde{\vec{H}}\}& \mathcal{R}\{\tilde{\vec{H}}\}\end{array}\right]\!,\eea
we can rewrite (\ref{complexmodel}) as a real-valued model
\be \label{realmodel} {\vec{y}}={\vec{H}}{\vec{x}} + {\vec{n}}\ee
where the $K\!\times\! K$ channel matrix $\vec{H}$ is known to the receiver, $\vec{x}\!=\![x_1 \, \ldots \, x_K]^\mathrm{T}$ contains PAM symbols from an alphabet $\mathcal{A}=\{\pm 1, \, \pm 3, \, \ldots \, ,\pm (\sqrt{M}-1)\}$, and $\vec{n}$ is random Gaussian noise with a covariance matrix $(N_0/2)\vec{I}$. As the transmit power depends on $M$, the signal-to-noise (SNR) is defined as
\bea \label{snr} \text{SNR}=2\mathbb{E}[|x_k|^2]/N_0. \eea 

The ZF detector is given by
\bea \hat{\bf x} = \mathcal{Q}_{\mathcal{A}}\left(\vec{H}^+ \vec{y}\right)\!,\eea
where $\mathcal{Q}_{\mathcal{A}}(\cdot)$ denotes entry-wise quantization to the nearest point in $\mathcal{A}$.
This can be slightly rewritten as
$$\hat{x}_k = \mathcal{Q}_{\mathcal{A}}(r_k),\quad 1 \leq k \leq K$$
with
\be \label{origZF} r_k=\boldsymbol{\delta}_k\vec{H}^+ \vec{y}\ee
and
$$\boldsymbol{\delta}_k = [\underbrace{0\; \ldots \; 0}_{k-1} \; 1 \; \underbrace{0\; \ldots \; 0}_{K-k}].$$
For later use, we note that we may just as well replace the \lq\lq{}1\rq\rq{} with any arbitrary scalar value and equivalently work with 
\bea r_k= \tau x_k + w_k,\eea
where $w_k$ is zero-mean complex Gaussian noise with  variance $\tau^2 (N_0/2)\|\boldsymbol{\delta}_k \vec{H}^+\|^2$.
Accordingly, the post-processing SNR, which is independent of $\tau$, becomes
\bea \label{psnr} \gamma_k = \frac{\text{SNR}}{\|\boldsymbol{\delta}_k \vec{H}^+\|^2},\eea
where  $\text{SNR}$ is defined in (\ref{snr}).

\section{Description of the Proposed MZF Detection}
A main issue with ZF is that $\|\boldsymbol{\delta}_k \vec{H}^+\|^2$ in (\ref{psnr}) is typically large and results in noise-enhancement. To combat that, we make use of the underlying idea of THP but apply it to detection, \textit{without any involvement of the transmitter.} 
We propose to replace (\ref{origZF}) with
\be\label{newZF} r_k=(\tau\boldsymbol{\delta}_k+\vec{q}_k)\vec{H}^+ \vec{y}\ee
where $\vec{q}_k \!=\! \left[q_{k1}, \, q_{k2}, \, \ldots \, q_{kK}\right]$ and $q_{k\ell}\!\in\! 2\mathbb{Z}$, i.e., the even integers. With that,
\be \label{withthat}
r_k= \tau x_k + \sum_{\ell=1}^{K} q_{k\ell} x_{\ell} +w_k,
\ee
and the post processing SNR  changes to 
\bea \label{psnrMZF} \gamma_k = \frac{\tau^2 \text{SNR}}{\|(\tau \boldsymbol{\delta}_k + \vec{q}_k)\vec{H}^+\|^2},\eea
rather than (\ref{psnr}).

Note that (\ref{withthat}) coincides with the received signal per user in a vector perturbation (VP) system \cite{HPS05, MJM11}. Therefore, further processing of (\ref{withthat})  follows the same steps as those for VP. Based on (\ref{withthat}), hard-output detection of $x_k$ from $r_k$ can be obtained based on the following property.
\begin{property}
Let 
$$y=z + \alpha\sum_{m=1}^{M} p_{m} b_{m},$$
where $\alpha\geq 1$, $|z|\!<\!2$, and $p_m, (b_m\!-\!1 )\!\in\! 2\mathbb{Z}$. Then,
{\setlength\arraycolsep{1pt} \bea \label{z1} z=\left\{\begin{array}{ll} (y \;\,\mathrm{mod} \;\,4\alpha)-2\alpha ,&\;\;\mathrm{if}\;\; \frac{1}{2}\sum\limits_{m=1}^M p_{m} \; \mathrm{is} \; \mathrm{odd},\\
\big((y+2\alpha) \;\,\mathrm{mod} \;\,4\alpha\big)-2\alpha,&\;\;\mathrm{otherwise}. \end{array}\right. \qquad \eea}
\end{property}
\begin{proof}
See Appendix A.
\end{proof}

Let us now consider $\alpha=1$ (other values are considered in Section \ref{exten1}). In view of Property 1, we see that  $q_{k\ell}$ and $x_{\ell}$ in (\ref{withthat}) qualify as $p_m$ and $b_m$. Further, from the condition $|z|<2$ in Property 1, we have that $\tau$ must be selected so that 
\bea \label{cond1} \tau\max_{a\in \mathcal{A}} |a| = \tau (\sqrt{M}-1)<2.\eea
To finalize the detector, we let
{\setlength\arraycolsep{1pt}\bea \label{zkdef} z_k =\left\{\begin{array}{ll} (r_k \;\,\mathrm{mod}\;\, 4) -2, & \;\;\mathrm{if}\;\;\frac{1}{2}\sum\limits_{\ell=1}^K q_{k\ell} \; \mathrm{is} \;\mathrm{odd}, \\
\big((r_k+2) \;\,\mathrm{mod}\;\, 4\big) -2, &\;\; \mathrm{otherwise},\end{array} \right. \quad\eea}
\hspace{-1.4mm}which can be expressed as 
$$z_k = \tau x_k+\tilde{w}_k,$$
where $\tilde{w}_k$ has a complicated distribution due to the modulus operation. The detected symbol $\hat{x}_k$ can now be obtained as
\bea \hat{x}_k = \mathcal{Q}_{\tau\mathcal{A}}(z_k),\eea
where the quantization is implemented on $\tau\mathcal{A}$, with the definition  $\tau \mathcal{A}=\{\tau x :x\in \mathcal{A}\}$.

We remark that, the choice $\tau\!=\!2/ (\sqrt{M}-1)$ is not suitable in (\ref{cond1}). This is so since if $\tau x_k\! +\! w_k\!=\!2\!+\!\epsilon$, for some small $\epsilon\!>\!0$, then $z_k\!=\!-2\!+\!\epsilon$. However, provided that $\tau\!\ll\!2/ (\sqrt{M}\!-\!1)$, such wrap seldomly happens at high SNR and $\tilde{w}_k \!=\! w_k$ with high probability. Observe that for constellation points $x_k$ with small magnitude, then $\tilde{w}_k = w_k$ with much higher probability than for constellation points $x_k$ of large magnitude. To ensure equal error probability for all constellation points, we design $\tau$ such that:

\textit{The distance from 2 to the largest constellation point in $\tau\mathcal{A}$ is half the distance between two points in $\tau\mathcal{A}$.} 

Following this rule results in 
\bea \tau = 2^{(1-\log_2(\sqrt{M}))},\eea
where with out loss of generality we assume that $\log_2(\sqrt{M})$ is an integer in the rest of the paper. With that, we have that $\tilde{w}_k$ is \lq\lq{}nearly\rq\rq{} Gaussian at high SNR; see VP-PAPER for further details.

To optimize the receiver, we should solve
\bea \label{opti} \vec{q}_k^{\mathrm{opt}}&=&\argmax_{\vec{q_k}} \gamma_k \nonumber \\
&=& \argmin_{\vec{q_k}} \|(\tau \boldsymbol{\delta}_k+\vec{q}_k)\vec{H}^+\|^2\eea
where elements of $\vec{q}_k$ are even integers. We rewrite (\ref{opti}) as
{\setlength\arraycolsep{1pt} \bea \label{Ab} \vec{q}_k^{\mathrm{opt}} &=& \argmin_{\vec{q}_k} \|\tau\boldsymbol{\delta}_k \vec{H}^+-\vec{q}_k(-\vec{H}^+)\|^2 \nonumber \\
&=&\argmin_{\vec{q}_k} \|\vec{b}_k-\vec{q}_k\vec{B}\|^2, \eea}
\hspace{-1.4mm}which is an instance of sphere detection over the integers \cite{AEZ02}, and
{\setlength\arraycolsep{1pt}\bea \vec{b}_k&=& \tau\boldsymbol{\delta}_k \vec{H}^+, \\
\vec{B}&=&-\vec{H}^+. \eea}
\hspace{2mm}Without any further extensions, pseudo-code for implementation of the  MZF detector is given in Algorithm 1. We remind the reader that the inputs $\vec{H}$ and $\vec{y}$ to the algorithm are assumed to be real-valued, while $M$ denotes the cardinality of the complex-valued QAM constellation. We provide a worked out  example of MZF detection in Appendix B to illustrate the process.

\subsection{Some Remarks on the MZF Detection}
With the principle of MZF detection introduced, we have a few important remarks as follows.
\begin{remark} 
The MZF is an extension of the ZF, where the latter is the special case of the former when 
\bea \label{badcond} \left\|\vec{q}_k^{\mathrm{opt}}-\boldsymbol{\delta}_k\odot\vec{q}_k^{\mathrm{opt}}\right\|^2\!=\!\vec{0}, \eea 
where $\odot$ denotes the Hadamard product. Therefore, from the perspective of post-processing SNR $\gamma_k$,  MZF is always superior to  ZF. 
\end{remark}
\begin{remark} Following Remark 1, when (\ref{badcond}) holds, the modulus operation degrades performance and should be removed.
\end{remark}
\begin{remark} In general the minimum value achieved by $\vec{q}_k^{\mathrm{opt}}$ in (\ref{Ab}) increases as $\tau$ decreases. That is, for alphabets with large cardinality, the gain of  MZF decreases.
\end{remark}
To resolve the issue in Remark 3 and to further improve performance, some useful extensions will be introduced in Sec. IV.

\begin{algorithm}[ht!]
\begin{algorithmic}[1]
\Function{$\hat{\vec{x}}=$ModularZF}{$\vec{H},\vec{y},M$}
\State $\tau = 2^{(1-\log_2(\sqrt{M}))}$
\State $\vec{B}=-\vec{H}^+$
\vspace*{-6mm}
   \NoNumber{\\\hrulefill}
\Statex $\quad\,\,\,$\textbf{Preprocessing for each coherence interval}
  \vspace*{-6mm} 
  \NoNumber{\\\hrulefill}
 \State{\textbf{for} $\, k = 1$ \textbf{to} $K$}
    \State $\quad\,\,\,$$\vec{b}_k = \tau\boldsymbol{\delta}_k \vec{H}^+$
    \State $\quad\,\,\,$$\mathrm{Solve:} \; \vec{q}^{\mathrm{opt}}_k=\argmin\limits_{\vec{q}_k} \|\vec{b}_k-\vec{q}_k\vec{B}\|^2$
		\State \textbf{end for}
   \vspace*{-6mm}
   \NoNumber{\\\hrulefill}

\Statex $\quad\,\,\,$\textbf{Executed for every channel observation}
    \vspace*{-6mm}
		\NoNumber{\\\hrulefill}
 \State{\textbf{for} $\, k = 1$ \textbf{to} $K$}
    \State $\quad\,\,\,\,\,r_k=(\tau \boldsymbol{\delta}_k+\vec{q}^{\mathrm{opt}}_k)\vec{H}^+\vec{y}$
     \State $\quad\,\,\,$ \textbf{if} $\left\|\vec{q}_k^{\mathrm{opt}}\!-\!\boldsymbol{\delta}_k\odot\vec{q}_k^{\mathrm{opt}}\right\|^2\!=\!\vec{0}$ \textbf{then}
     			   \State $\quad\quad\quad\,\,\,  z_k =\vec{b}_k\vec{y}$
     			       \State \quad\,\,\, \textbf{else} 
	   \State $\quad\quad\quad\,\,\,$ \textbf{if} $\frac{1}{2}\sum_{\ell=1}^K q^{\mathrm{opt}}_{k\ell} \;\; \mathrm{is} \;\; \mathrm{odd}$ \textbf{then}
	      \State $\quad\quad\quad\,\,\,\quad\,\,\,  z_k = (r_k \;\,\mathrm{mod}\;\, 4) -2$

		  \State $\quad\quad\quad\,\,\,$ \textbf{else}
		     \State $\quad\quad\quad\,\,\,\quad\,\,\,  z_k = ((r_k+2) \;\,\mathrm{mod}\;\, 4) -2$
	  \State $\quad\quad\quad\,\,\,$ \textbf{end if}
	       			       \State \quad\,\,\, \textbf{end if} 
		\State $\quad\,\,\,\hat{x}_k = \mathcal{Q}_{\tau\mathcal{A}}(z_k)$
		\State \textbf{end for}

\EndFunction
\end{algorithmic}
\caption{\label{alg:1} Modulus Zero-Forcing (MZF) Algorithm \newline \hspace*{18mm} $\vec{H}$ is $K\times K$ real-valued \newline \hspace*{18mm} $\vec{y}$ is $K\times 1$ real-valued \newline \hspace*{18mm} $M$ is cardinality of QAM constellation}
\end{algorithm}

\subsection{Discussion on the Diversity-Multiplexing Trade-off (DMT)}
In the designs of IF receivers \cite{ZG14, CJ17, OE10}, the target is to directly optimize
\bea \label{optIF} \vec{p}_k^{\mathrm{opt}}= \argmin_{\vec{p_k}} \|\vec{p}_k\vec{H}^+\|^2,\eea
given the constraint that $\vec{P}\!\in\!\mathbb{Z}^{K\times K}$ is full-rank\footnote{The rationale behind is that, in order to have $\vec{P}$ mod $p$ to be full-rank over $\mathbb{Z}_p$ (to recover the codewords from their linear combinations after decoding on $\mathbb{Z}_p$), it suffices to check whether $\vec{P}$ is full-rank over $\mathbb{R}$, given that the magnitudes of the elements of $\vec{P}$ are upper-bounded by a constant \cite[Th. 11]{NG11}.} over $\mathbb{R}$, where $\vec{P}$ it the matrix comprises all vectors $\vec{p}_k$. Comparing to (\ref{opti}), it can be seen that, with MZF we constrain $\vec{p}_k$ in (\ref{optIF}) to be
\bea \label{pk} \vec{p}_k=  \boldsymbol{\delta}_k+\frac{1}{\tau}\vec{q}_k,\eea
where $\vec{q}_k$ comprises even integers. With (\ref{pk}) it holds that
\bea \vec{P}=\vec{I}+\frac{1}{\tau}\vec{Q}, \eea
where $\tau\!<\!1$ and the elements of $\vec{Q}$ are even integers. Therefore, with MZF, the degrees-of-freedom (DoFs) in designing $\vec{P}$ is less than that of an IF receiver\footnote{However, with MZF matrix $\vec{P}$ is not required to be full-rank.}. Nevertheless, we have the following Property 2 (which is an extension of \cite[Theorem 5]{ZG14}) that shows that, the MZF also achieves the optimal DMT \cite{ZT03} as IF receiver does. Therefore, from an information-theoretic perspective, the MZF detection does not scarifies much performance compared to an IF receiver, while the latter one has much higher encoding/decoding complexity.
 
\begin{property}
For a complex-valued MIMO channel with $K$ transmit-antennas, $N\!\geq\!K$ receive-antennas, and independent and identically distributed (i.i.d.) Rayleigh fading, the achievable DMT with the MZF
detection\footnote{In this case, the size of $\vec{H}$ corresponding to the real-valued model in (\ref{realmodel}) is $2N\!\times\!2K$.} is
\bea d_{\mathrm{MZF}}=N\left(1-\frac{r}{K}\right),\; 0\leq r\leq K. \eea
\end{property}
\begin{proof}
See Appendix C.
\end{proof}

\section{Extensions of the MZF Detection}
In this section we introduce some extensions to the basic MZF detector for further improving its  performance. While Extension 1 and 4 are generalizations of the basic algorithm, Extension 2 is to resolve the issue mentioned for large cardinality alphabets  and improve the performance for weak bit-layers, while Extension 3 is a decision feedback version of Extension 2.

\subsection{Extension 1: A scaled modulus} \label{exten1}
This first extension arises from a slight relaxation of $\alpha\!=\!1$ in the MZF detector. From Property 1, we can replace (\ref{zkdef}) as
{\setlength\arraycolsep{0pt}\bea \label{zkdef1} z_k =\left\{\begin{array}{ll} (r_k \;\,\mathrm{mod}\;\, 4\alpha) -2\alpha, & \;\;\mathrm{if}\;\;\frac{1}{2}\sum\limits_{\ell=1}^K q_{k\ell} \; \mathrm{is} \;\mathrm{odd}, \\
\big((r_k+2\alpha) \;\,\mathrm{mod}\;\, 4\alpha\big) -2\alpha, &\;\; \mathrm{otherwise}.\end{array} \right. \qquad\eea}
\hspace{-1.4mm}This requires us to optimize, instead of (\ref{opti}),
 \bea \label{opti3} (\vec{q}_k^{\mathrm{opt}},\alpha^{\mathrm{opt}})= \argmin_{|\alpha| \geq 1,\; \vec{q}_k} \|(\tau \boldsymbol{\delta}_k+\alpha \vec{q}_k)\vec{H}^+\|^2.\eea

Solving (\ref{opti3}) is harder than solving (\ref{opti}) since it can be regarded as an instance of non-coherent sphere detection. Instead, we solve (\ref{opti}) first, and then plug the optimal solution into (\ref{opti3}) and solve for the optimal $\alpha$. That is, $\vec{q}_k^{\mathrm{opt}}$ is obtained with (\ref{opti}), and
\bea \alpha^{\mathrm{opt}}= \arg \min_{|\alpha|\geq 1} \|(\tau \boldsymbol{\delta}_k+\alpha\vec{q}_k^{\mathrm{opt}})\vec{H}^+\|^2 ,\eea
where we sligthly abused notation since the pair $(\vec{q}_k^{\mathrm{opt}},\alpha^{\mathrm{opt}})$ is in general not \textit{jointly} optimal in the sense of (\ref{opti3}). Although Extension 3 is intuitive, the gain seems marginal according to numerical results.

\subsection{Extension2: Bit-wise Modulus Zero-Forcing} \label{ext2}
An underlying assumption of this extension is that the bit-mapping to the symbols in $\mathcal{A}$ is such that the constellation has an additive structure. By this we mean that a PAM symbol $x_k$ should be of the form
\bea x_k = \sum_{b=1}^{\log_2(\sqrt{M})}\!\!u_{kb} 2^{b-1}, \eea
where $u_{kb}\in\{\pm 1\}$ correspond to information bits.

Using Algorithm 1, the bits $u_{kb}$ are determined by the output $\hat{x}_k\! =\! \mathcal{Q}_{\tau\mathcal{A}}(z_k)$, with setting $\tau \!=\! 2^{(1-\log_2(\sqrt{M}))}$. As $M$ increases, $\tau$ decreases and so are the gains of MZF detection. To resolve this for large values of $M$, we extend the symbol-based MZF detector in Algorithm 1 to a bit-wise version.

Note that we can rewrite (\ref{withthat}) as
\bea \label{withthat2} 
r_k = \tau \sum_{b=1}^{\log_2(\sqrt{M})}u_{kb} 2^{b-1} +\sum_{\ell=1}^K q_{k\ell} x_{\ell} + w_k.
\eea
Assuming that we are interested in the $n$-th bit $u_{kn}$, we let
\bea \tilde{x}_k = \sum_{b=1}^{n}u_{kb} 2^{b-1},\eea
which belongs to a $2^n$-PAM alphabet. Setting $\tau(n)\!=\!2^{1-n}$ in (\ref{withthat2}) yields
\be
r_k = 2^{1-n}\tilde{x}_k +\sum_{b=n+1}^{\log_2(\sqrt{M})}u_{kb} 2^{b-n} +\sum_{\ell=1}^K q_{k\ell} x_{\ell} + w_k.
\ee
It can be easily seen that $\frac{1}{2}\sum_{b=n+1}^{\log_2(\sqrt{M})}u_{kb} 2^{b-n}$ is an odd integer so it qualifies as a valid value of $q_{k\ell}$, and $u_{kn}$ can be detected as
\bea \label{ukn} \hat{u}_{kn} = \mathrm{sign}(z_k).\eea

Therefore, for each bit-layer, a different value of $\tau$ is used and only a sign operation is needed for detection. Extension 2 has a complexity increment over Algorithm 1 since an optimization to find  $\vec{q}_k$ is needed for each bit-layer. 

Note that, according to Remark 2, when detecting the last bit-layer and if (\ref{badcond}) holds, the ZF estimate shall be used for detection, while for detecting the other layers, modulus operations are still needed to cancel the transmitted bits corresponding to higher bit-layers. Psuedo-code for the MZF with this extension  is summarized in Algorithm 2.

\begin{algorithm}[ht!]
\begin{algorithmic}[1]
\Function{$\hat{\vec{x}}=$ModularZFExt2}{$\vec{H},\vec{y},M$}
\State{$\quad\,\,\,N =\log2(\sqrt{M})$}
\State $\quad\,\,\,\vec{B}=-\vec{H}^+$
 \State $\quad\,\,\,\textbf{for}$ \, $n = 1$ \textbf{to} $N$
  \State $\quad\quad\quad\,\,\ \tau(n) = 2^{1-n}$
  \State $\quad\,\,\,\textbf{end for}$
\vspace*{-6mm}
   \NoNumber{\\\hrulefill}
\Statex $\quad\,\,\,\quad\,\,\,$\textbf{Preprocessing for each coherence interval}
  \vspace*{-6mm} 
  \NoNumber{\\\hrulefill}
 \State{$\quad\,\,\,\textbf{for}\,$  $k = 1$ \textbf{to} $K$}
  \State $\quad\quad\quad\,\vec{b}_{k} =\boldsymbol{\delta}_k \vec{H}^+$
 \State $\quad\quad\quad\,\textbf{for}$ \, $n = 1$ \textbf{to} $N$
    \State $\quad\quad\quad\quad\,\,\,$$\vec{b}_{kn} = \tau(n)\vec{b}_{k}$
    \State $\quad\quad\quad\quad\,\,\,$$\mathrm{Solve:} \; \vec{q}^{\mathrm{opt}}_{k,n}=\argmin\limits_{\vec{q}_k} \|\vec{b}_{kn}-\vec{q}_k\vec{B}\|^2$
    \State\quad\quad\quad\,\textbf{end for}
		\State $\quad\,\,\,$\textbf{end for}
   \vspace*{-6mm}
   \NoNumber{\\\hrulefill}

\Statex $\quad\,\,\,\quad\,\,\,$\textbf{Executed for every channel observation}
    \vspace*{-6mm}
		\NoNumber{\\\hrulefill}
 \State{$\quad\,\,\,$\textbf{for} $\, k = 1$ \textbf{to} $K$}
  \State{$\quad\quad\quad\,$\textbf{for} $\, n = 1$ \textbf{to} $N$}
       \State \quad\quad\quad\quad\,\textbf{if} $n\!=\!N$ \textbf{and} $\left\|\vec{q}_k^{\mathrm{opt}}\!-\!\boldsymbol{\delta}_k\odot\vec{q}_k^{\mathrm{opt}}\right\|^2\!=\!\vec{0}$,  \textbf{then}
              \State \quad\quad\quad\quad\quad\,\,\,\,\,$z_k=\vec{b}_{k}\vec{y}$
                  \State \quad\quad\quad\quad\,\textbf{else}
    \State $\quad\quad\quad\quad\quad\,\,\,\,\,r_k=(\tau(n) \boldsymbol{\delta}_k+\vec{q}^{\mathrm{opt}}_k)\vec{H}^+\vec{y}$
	   \State $\quad\quad\quad\quad\quad\,\,\,\,\, z_k = (r_k \;\,\mathrm{mod}\;\, 4) -2$
	           \State\quad \quad\quad\quad\,\textbf{end if}
		\State $\quad\quad\quad\quad\quad\quad\,\,\,\,\, \hat{u}_{kn} = \sign(z_k)$
		  \State{$\quad\quad\quad\,$\textbf{end for}}
				\State $\quad\,\,\,$\textbf{end for}
\EndFunction
\end{algorithmic}
\caption{\label{alg:2} MZF Algorithm with Extension 2 \newline \hspace*{18mm} $\vec{H}$ is $K\times K$ real-valued \newline \hspace*{18mm} $\vec{y}$ is $K\times 1$ real-valued \newline \hspace*{18mm} $M$ is cardinality of QAM constellation}
\end{algorithm}

\subsection{Extension 3: A decision feedback version of Extension 2}
An obstacle with Extension 2 is that $\tau(n)$ decreases as $n$ grows, and as previously mentioned, performance deteriorates. Small values of $n$ correspond to weak bit-layers, and large $n$ correspond to strong bit-layers. Thus, with Extension 2, predominantly the weak bit-layers can gain by the MZF, while the gain could be minuscule for strong bit-layers. 
A gain for weak bit-layers is important since it is typically these bit-layers that limit ultimate performance. However, we can also harvest a gain for strong bit-layers via a decision feedback mechanism. To prevent error propagation in decision feedback equalization, strong bits are typically detected first and then canceled. That option is not available for MZF, rather we detect the weakest bit-layer first and then move on to stronger ones.

The method works as follows. First set $n\!=\!1$ and follow  Extension 2 verbatim to obtain $\hat{\vec{u}}_1 \!=\! [\hat{u}_{11} \, \ldots\, \hat{u}_{K1}]^{\mathrm{T}}$. For notational convenience, define $\vec{y}_1\!=\!\vec{y}$. Now construct
\bea \vec{y}_2 = \frac{1}{2}\left(\vec{y}_1-\vec{H}\hat{\vec{u}}_1\right). \eea
Provided that $\hat{\vec{u}}_1$ is correct, $\vec{y}_2$ is described with the same MIMO channel as $\vec{y}_1$, but with $\sqrt{M}/2$-PAM rather than $\sqrt{M}$-PAM inputs. Next, move on to $n\!=\!2$ and keep $\tau(2)\!=\!1$. Since nor the value of $\tau(2)$ neither the channel $\vec{H}$ has changed, the optimal vector $\vec{q}_k$ for $n\!=\!2$ coincides with that already found for $n\!=\!1$. We then have that for $\vec{y}_2$ 
\be \label{ext33}
r_k =  u_{k2} + \sum_{b=3}^{\log_2(\sqrt{M})}u_{kb} 2^{b-2} +\sum_{\ell=1}^K q_{k\ell} \frac{1}{2}(x_{\ell}-\hat{u}_{k1}) + w_k,
\ee
and $\hat{u}_{k2}$ is obtained by taking the sign of $z_k$ as in (\ref{ukn}). We proceed by 
\bea \vec{y}_3 = \frac{1}{2}\left(\vec{y}_2-\vec{H}\hat{\vec{u}}_2\right),\eea
and continue the process until all bit-layers have been detected.

Similarly, according to Remark 2, whenever (\ref{badcond}) holds, the ZF estimate shall be used. Pseudo-code is provided in Algorithm 3.

Extension 3 is similar to Extension 2 in the sense that, the detection for all bit-layers only needs to take the signs of $z_k$ as in (\ref{ukn}), but it has less complexity since only one optimization of (\ref{opti}) is needed which is shared for all bit-layers. A drawback with Extension 3 is that, as for all decision-feedback based detectors, the processing of the bit-layers cannot be parallelized, which is however, possible with Extension 2. Another drawback is potential  error-propagation at low SNRs.

\subsection{Extension 4: Replacing ZF by LMMSE}
So far we have introduced  modulus  arithmetic detection using ZF, however, $\vec{H}^{+}$ can also be replaced by other linear detectors\footnote{This is also known as regularized perturbation in VP \cite{HPS05}.} such as LMMSE, which sets
\bea \label{le} \vec{H}^{+}=\vec{H}\rmh\left(\vec{H}\vec{H}\rmh+N_0\vec{I}\right)^{-1}\!. \eea
In vector form, and with the introduction of a  matrix $\vec{T}$, the received signal after equalization is
\bea  \vec{T}\vec{H}^{+}\vec{y}=\vec{T}\vec{x}+\vec{T}\left(\vec{H}^{+}\vec{H}-\vec{I}\right)\vec{x}+\vec{T}\vec{H}^{+}\vec{n}, \notag \eea
where $\vec{T}\!=\!(\tau\vec{I}\!+\!\vec{Q})$. The target of optimizing $\vec{q}_k$ in this case, is to minimize the interference plus noise power that equals
\bea  \tilde{\vec{q}}_k=\argmin_{\vec{q}_k}\left\|(\tau\vec{\delta}_k+\vec{q}_k)\vec{E}\right\|^2, \eea
where
\bea \vec{E}=\left[\vec{H}^{+}\vec{H}-\vec{I}, \; N_0\vec{H}^{+}\right]\!. \eea

Note that, when $\vec{H}^{+}$ is the pseudo-inverse of $\vec{H}$ such as with ZF, $\vec{E}$ degrades to $\vec{B}$, which shows the generalization of the MZF concept. The reason for introducing Extension 4 is that, the ZF detector is suboptimal to LMMSE at low SNRs, in which case the modulus operation based on LMMSE can improve  performance. Therefore, it is beneficial to use LMMSE instead of ZF. Since only $\vec{H}^{+}$ is replaced by LMMSE equalization in Extension 4, all Algorithms 1-3 still apply verbatim.

There are also many other possible variations of the MZF, but which we will not pursue any further. Next we put an interest on comparing the MZF to a traditional LAR detector. The reason is that, solving (\ref{opti}) involves significant complexity, and we put forth an approximated solution based on LR with less computational efforts.

\begin{algorithm}[ht!]
\begin{algorithmic}[1]
\Function{$\hat{\vec{x}}=$ModularZFExt3}{$\vec{H},\vec{y},M$}
\State $\quad\,\,\,\tau=1$
\State{$\quad\,\,\,N =\log_2(\sqrt{M})$}
\State $\vec{B}=-\vec{H}^+$
\vspace*{-6mm}
   \NoNumber{\\\hrulefill}
\Statex $\quad\,\,\,\quad\,\,\,$\textbf{Preprocessing for each coherence interval}
  \vspace*{-6mm} 
  \NoNumber{\\\hrulefill}
 \State{$\quad\,\,\,$\textbf{for} $\, k = 1$ \textbf{to} $K$}
    \State $\quad\quad\quad\quad\,\,\,$$\vec{b}_k = \tau\boldsymbol{\delta}_k \vec{H}^+$
    \State $\quad\quad\quad\quad\,\,\,$$\mathrm{Solve:} \; \vec{q}^{\mathrm{opt}}_{k}=\argmin\limits_{\vec{q_k}} \|\vec{b}_k-\vec{q}_k\vec{B}\|^2$
		\State $\quad\,\,\,$\textbf{end for}
   \vspace*{-6mm}
   \NoNumber{\\\hrulefill}

\Statex $\quad\,\,\,\quad\,\,\,$\textbf{Executed for every channel observation}
    \vspace*{-6mm}
		\NoNumber{\\\hrulefill}
        		            		    \State{$\quad\,\,\,\hat{\vec{y}} =\vec{y}$}
 \State{$\quad\,\,\,$\textbf{for} $\, n = 1$ \textbf{to} $N$}
         		    \State{$\quad\quad\quad\,\hat{\vec{u}}_n =0$}
  \State{$\quad\quad\quad\,$\textbf{for} $\, k = 1$ \textbf{to} $K$}
         \State \quad\quad\quad\quad\,\textbf{if} $\left\|\vec{q}_k^{\mathrm{opt}}\!-\!\boldsymbol{\delta}_k\odot\vec{q}_k^{\mathrm{opt}}\right\|^2\!=\!\vec{0}$,  \textbf{then}
         \State $\quad\quad\quad\quad\quad\,\,\,\,\,z_k=\tau \boldsymbol{\delta}_k\vec{H}^+\hat{\vec{y}}$    
                 \State\quad \quad\quad\quad\,\textbf{else}
    \State $\quad\quad\quad\quad\quad\,\,\,\,\,r_k=(\tau \boldsymbol{\delta}_k+\vec{q}^{\mathrm{opt}}_k)\vec{H}^+\hat{\vec{y}}$
	   \State $\quad\quad\quad\quad\quad\,\,\,\,\, z_k = (r_k \;\,\mathrm{mod}\;\, 4) -2$
	            \State \quad\quad\quad\quad\,\textbf{end if}
		\State $\quad\quad\quad\quad\,\hat{u}_{kn} = \sign(z_k)$
		  \State{$\quad\quad\quad\,$\textbf{end for}}
		  		   \State{\quad\quad\quad\,$\hat{\vec{u}}_n \!=\! [\hat{u}_{1n} \, \hat{u}_{2n} \, \ldots\, \hat{u}_{Kn}]^{\mathrm{T}}$}
		  		       \State $\quad\quad\quad\,\hat{\vec{y}}=(\hat{\vec{y}}-\vec{H}\hat{\vec{u}_n})/2$
				\State $\quad\,\,\,$\textbf{end for}
\EndFunction
\end{algorithmic}
\caption{\label{alg:3} MZF Algorithm with Extension 3 \newline \hspace*{18mm} $\vec{H}$ is $K\times K$ real-valued \newline \hspace*{18mm} $\vec{y}$ is $K\times 1$ real-valued \newline \hspace*{18mm} $M$ is cardinality of QAM constellation}
\end{algorithm}

\section{A Solution Based on, and a Comparison to, Lattice Reduction}
Except for approximately solving (\ref{opti}) with LR, another reason for comparing MZF with LR detection is that, the obtained MZF allows for a direct comparison to LAR detectors. In LAR as well as the MZF, the most burdening task is to execute the LLL algorithm (or other similar algorithms), thus the complexities of LAR and MZF become virtually identical. As we will demonstrate, the detection-performance of MZF is superior in some cases. 
\subsection{A quick review of LAR}
Given (\ref{realmodel}), LAR starts by performing the LLL algorithm on $\vec{H}$, so that we obtain $\bar{\vec{H}} = \vec{H}\vec{T}$ where $\vec{T}$ is unimodular and $\bar{\vec{H}}$ is nearly orthogonal. With $\vec{z} = \vec{T}^{-1}\vec{x}$ we have
\bea \vec{y} =\bar{\vec{H}}\vec{z} + \vec{n}. \eea
Performing ZF based on $\bar{\vec{H}}$ and quantizing to the nearest integers gives
\be \label{quantLAR} \hat{\vec{z}} = \mathcal{Q}_{\mathbb{Z}}(\bar{\vec{H}}^{-1}\vec{y})\ee
from which one can obtain
$$\hat{\vec{x}} = \mathcal{Q}_{\mathcal{A}}(\vec{T}\hat{\vec{z}}).$$
Clearly, once $\bar{\vec{H}}$ has been established, the remaining steps are of miniscule complexity.

At this point, a reasonable question is, what the relation between LAR and MZF is, and whether they are equivalent? The answers to these questions are that, they are closely related, but not equivalent.
Prior to quantization in (\ref{quantLAR}), we can write
{\setlength\arraycolsep{2pt}\bea \label{c1} \vec{r}&=& \bar{\vec{H}}^{-1}\vec{y}  \nonumber \\
&=& \vec{T}^{-1}\vec{x} + \vec{w}. \eea}
\hspace{-1.4mm}Since $\vec{T}$ is unimodular, so is $\vec{T}^{-1}$. 

On the other hand, written in vector form, (\ref{newZF}) equals
{\setlength\arraycolsep{2pt}\bea \label{c2} \vec{r} &=&\left(\tau\vec{I} + \vec{Q}\right) \vec{H}^{-1}\vec{y} \nonumber \\
&=& \left(\tau\vec{I} + \vec{Q}\right)\vec{x} + \vec{w}. \eea}
\hspace{-1.4mm}Comparing (\ref{c1}) and (\ref{c2}) with $\tau\!=\!1$, we see that in both cases $\vec{r}$ equals an integer-valued matrix multiplied with the data symbols, plus noise. However, the matrix $\vec{T}^{-1}$ in (\ref{c1}) has no particular structure (besides being unimodular) so the modulus operation in (\ref{zkdef}) is not available. This makes LAR, i.e., (\ref{c1}) and MZF, i.e., (\ref{c2}) fundamentally different, as the structure of (\ref{c1}) requires further processing in the form of (\ref{quantLAR}) while (\ref{c2}) allows for further processing via (\ref{zkdef}).

\subsection{An approximate solution to (\ref{opti}) based on LLL}
In (\ref{opti}) we have the following problem to solve
\bea \label{opti4}\vec{q}^{\mathrm{opt}}=\arg \min_{\vec{q}} \|\vec{b}-\vec{q} \vec{B}\|^2,\eea
where we removed the subscript $k$, and the vectors are row-vectors. Perform the LLL algorithm to $\vec{B}^{\mathrm T}$ so that we have $$\bar{\vec{B}}=\vec{B}^{\mathrm T}\vec{T}.$$ Since $\vec{B} \!=\! -\vec{H}^+$ the LLL algorithm needs, similar to LAR, to be executed only once per coherence interval. We can now proceed as in the LAR case, 
$$\hat{\vec{z}} = \mathcal{Q}_{\mathbb{Z}}(\bar{\vec{B}}^{-1}\vec{b}^{\mathrm T})$$
followed by
\bea \vec{q}^{\mathrm{opt}} = \left[\mathcal{Q}_{2\mathbb{Z}}(\vec{T}\hat{\vec{z}})\right]^{\mathrm T}.\eea

Note that, the optimization (\ref{opti4}) itself is also a MIMO detection problem (but only needs to run once per coherence-interval), therefore, there are also other low-complexity suboptimal algorithms to solve (\ref{opti4}), such as using ZF or partial marginalization \cite{HR17j1}. In the simulations, we will focus on the optimal SD and the suboptimal LLL  solutions for (\ref{opti4}), respectively.

\section{Numerical Results} \label{numres}
In this section, we show some numerical results of the proposed MZF detector, as well as its extensions. In all tests, we test with $K\!\times\!K$ real-valued MIMO channels (each element is an independent and identically distributed Gaussian variable with a zero-mean and unit-variance) with $\sqrt{M}$-PAM modulated symbols that are transfered from $K/2\!\times\!K/2$ complex-valued MIMO channels  and $M$-QAM modulated symbols. We simulate 50,000 channel realizations for each of the tests.

\subsection{SINR improvements}
\begin{figure}[t]
\begin{center}
\scalebox{.42}{\includegraphics{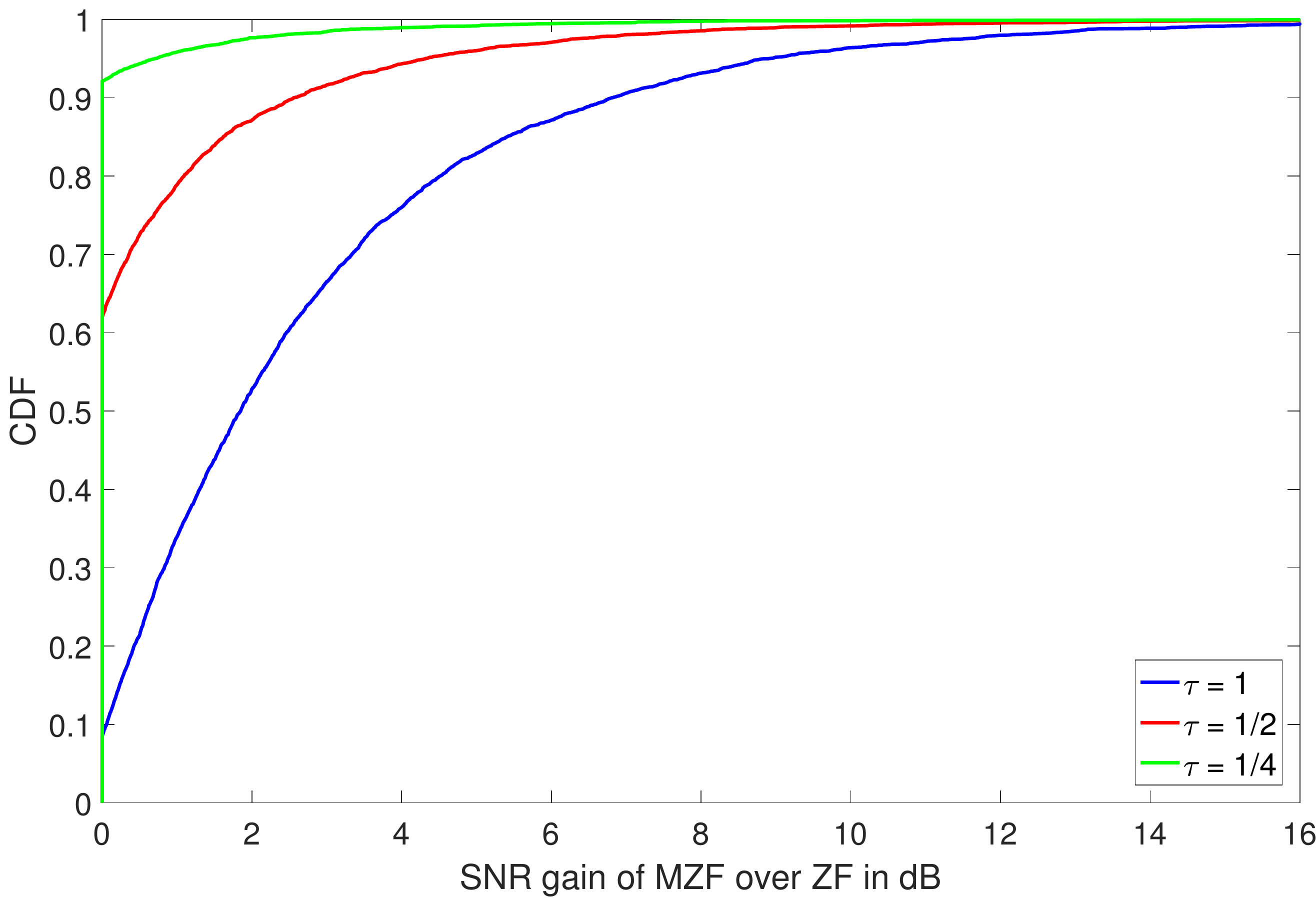}}
\vspace*{-3mm}
\caption{\label{fig2} SNR gains under real-valued $12\!\times\!12$ MIMO with 2-PAM, 4-PAM and 8-PAM modulations, respectively.}
\vspace*{-4mm}
\end{center}
\end{figure}

In Fig. \ref{fig2} we show the post-processing SNR improvements with the MZF detector using Algorithm 1, and compare to a traditional ZF detector with different PAM modulations (i.e., $\tau$ values). As can be seen, the SNRs are greatly improved, especially for low-order modulations (or the weak bit-layers of high-order modulations with Extension 2). When $\tau$ decreases, the gains become smaller. We also test the MZF with Extension 1, where we can observe only marginal gains (not shown in Fig. \ref{fig2}). Therefore, in the remaining tests we  set $\alpha\!=\!1$.

\begin{figure}
\begin{center}
\scalebox{.42}{\includegraphics{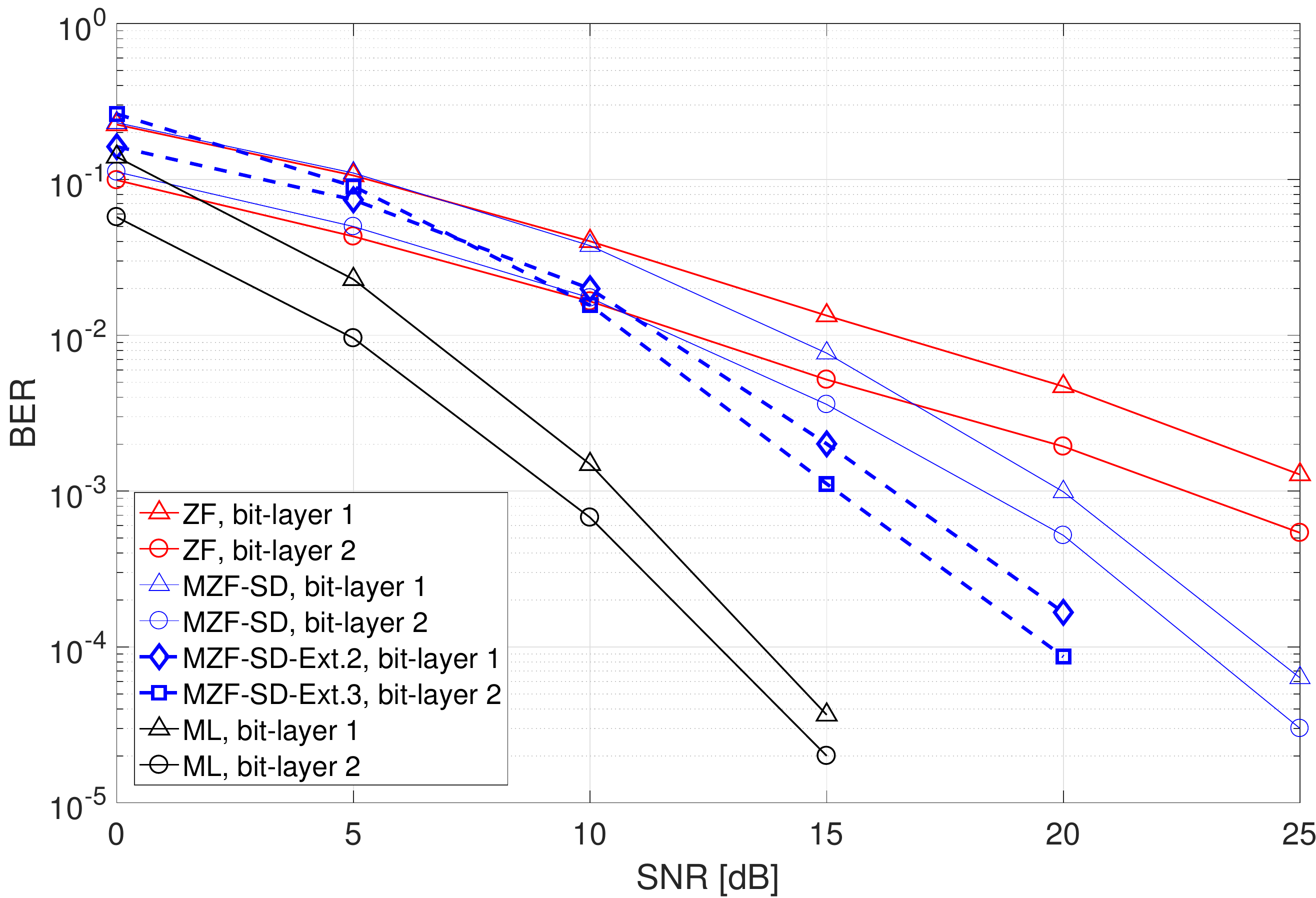}}
\vspace*{-3mm}
\caption{\label{fig3} Uncoded BER under real-valued $6\!\times\!6$ MIMO with 4-PAM modulation.}
\vspace*{-6mm}
\end{center}
\end{figure}

\begin{figure}[t]
\begin{center}
\scalebox{.42}{\includegraphics{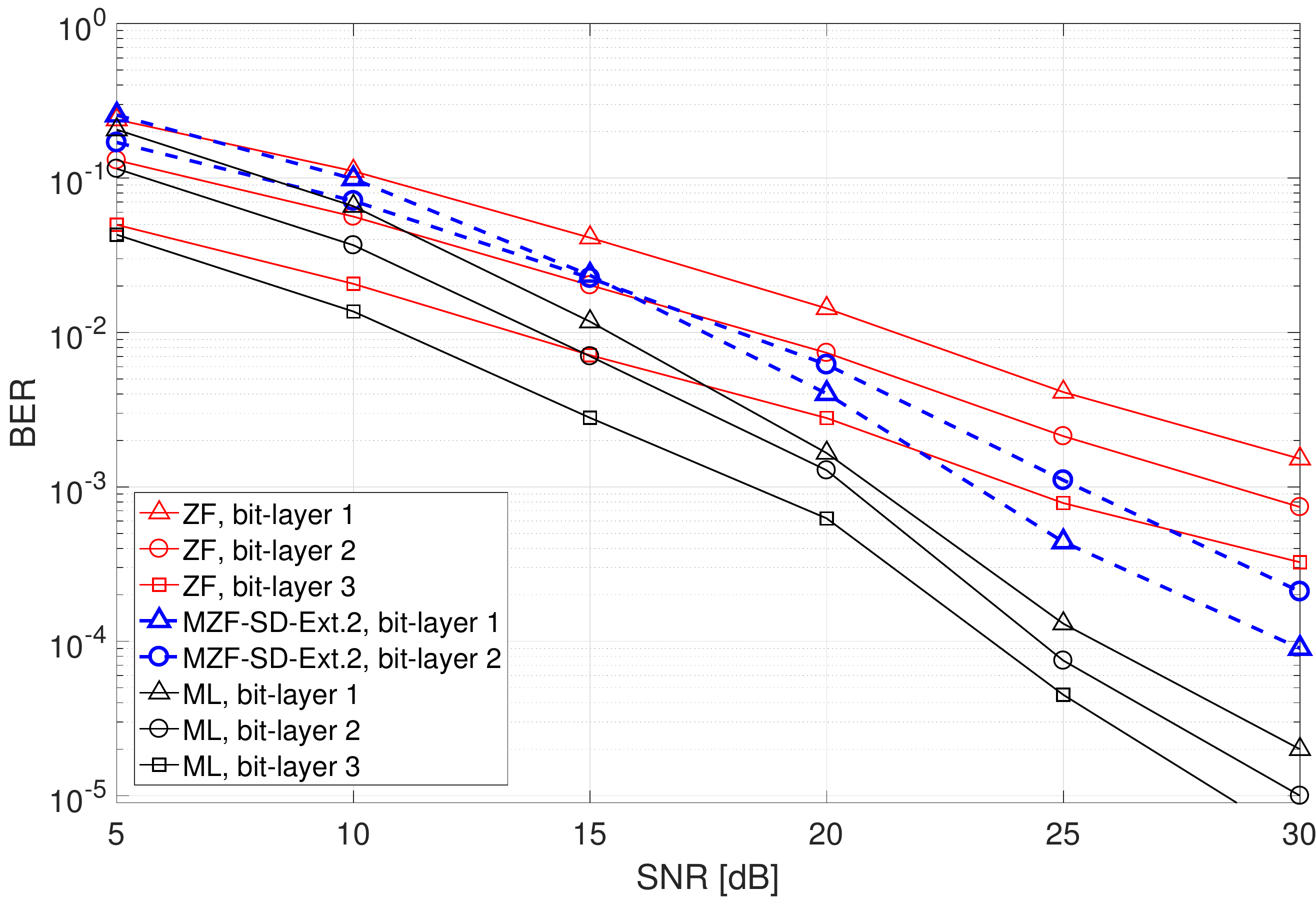}}
\vspace*{-3mm}
\caption{\label{fig4} Uncoded BER under real-valued $4\!\times\!4$ MIMO with 8-PAM modulation.}
\vspace*{-4mm}
\end{center}
\end{figure}

\begin{figure}
\begin{center}
\scalebox{.42}{\includegraphics{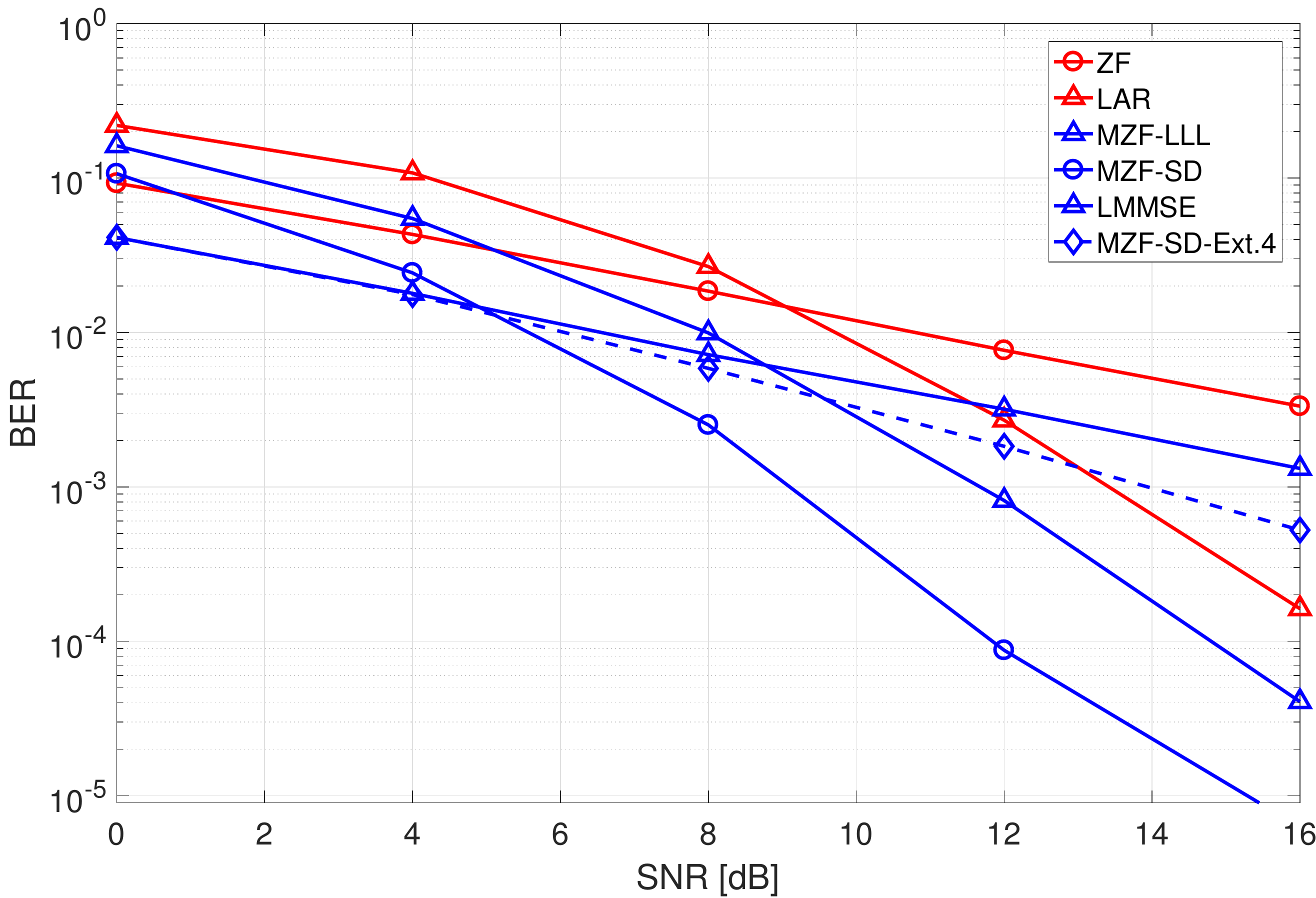}}
\vspace*{-3mm}
\caption{\label{fig5} Uncoded BER under real-valued $8\!\times\!8$ MIMO with 2-PAM modulation.}
\vspace*{-6mm}
\end{center}
\end{figure}

\subsection{Uncoded bit-error-rate (BER)}

Next we show  uncoded BER performance. In Fig. \ref{fig3} we compare MZF with ZF and ML under $6\!\times\!6$ MIMO with 4-PAM modulation. The MZF uses SD to find the optimal $\vec{q}_k$. As can be seen, the MZF without extensions outperforms the ZF with more than 2 dB at 0.1\% BER. With Extension 2, the BER of the first bit-layer (weaker layer) is greatly improved by more than 4 dB at 0.1\% BER and outperforms the second bit-layer, which justifies the application of Extension 3. With Extension 3, where the feedbacks of the first bit-layer are used, the BER of the second bit-layer is also improved by more than 3 dB at 0.1\% BER compared to  MZF with Extension 2. The gaps between ZF and ML are significantly reduced by the MZF, and the slopes of BER with MZF are also much steeper than the ZF, and close to those with the ML. However, as also can be observed, the MZF has only marginal gains at low SNRs, and the decision-feedback approach performs even worse due to inaccurate feedbacks. This obstacle can be relieved by using LMMSE based approaches, i.e., Extension 4.

In Fig. \ref{fig4} we repeat the tests in Fig. \ref{fig3} under $4\!\times\!4$ MIMO with 8-PAM modulation, that is, three bit-layers are considered. The MZF with Extension 2 using SD is compared to ZF and ML. As already shown in Fig. \ref{fig2}, setting $\tau\!=\!1/4$ for detecting the third-layer (strongest layer) only has small gains, and the BER performance is also close to ZF and therefore are not shown in Fig. \ref{fig4}. Nevertheless, the BER of the first and second bit-layers are significantly boosted by MZF. As can be seen, the MZF performs around 3 dB better than the ZF at 0.1\% BER for the second bit-layer, and 7 dB better for the first bit-layer. Since the weakest bit-layer usually has a stronger impact on the decoding performance, the gains for the first bit-layer is of importance.

\subsection{Comparison with LAR}

In Fig. \ref{fig5} we compare the MZF with the LAR under $8\!\times\!8$ MIMO with 2-PAM modulation. The MZF uses both SD and the approximate LLL method to find the optimal $\vec{q}_k$. As can be seen,  MZF outperforms  LAR with more than 1.5 dB at 0.1\% BER, with a similar complexity for running LLL algorithm for LR. 

Moreover, with Extension 4 (the LMMSE based detection), the BERs at low SNRs are also improved with the MZF which is inferior to the original ZF based MZF at high SNRs. Nevertheless, with Extension 4, the MZF is 4 dB better than a normal ZF, and more than 2dB better than a normal LMMSE detector at 0.1\% BER. Another observation is that, the SD based MZF is more than 2 dB better than the LLL based MZF, which shows that optimal selection of $\vec{q}_k$ is important.

%


\section{Summary}
We have proposed a novel modulus arithmetic  based zero-forcing (MZF) detector for multi-input multi-output (MIMO) channels, with a number of possible extensions of the basic algorithm. The MZF detector shows significant gains in terms of post-processing signal-to-noise-ratio (SNR) and bit-error-rate (BER) compared to traditional linear detectors, at medium and high SNR scenarios and in particular for weak bit-layers. At low SNRs and with large cardinalities of the input alphabet, we have provided several possible extensions to improve the performance of the MZF. Finding the optimal modulus matrix itself is a burdening MIMO detection problem, but it needs to be done only once per a coherence-interval of the MIMO channel using sphere-decoding (SD) or other suboptimal algorithms. In particular, with a similar complexity,  MZF with lattice-reduction (LR) based approaches outperforms the traditional lattice-aided-reduction (LAR) detector, which justifies its potential in MIMO detection.

\section*{Appendix A: Proof of Property 1}
Since $p_m, (b_m\!-\!1 )\!\in\! 2\mathbb{Z}$, we let $p_m\!=\!2\tilde{p}_m$ and $b_m\!=\!2\tilde{b}_m\!+\!1$, where $\tilde{p}_m, \tilde{b}_m\!\in\! \mathbb{Z}$. Then,
{\setlength\arraycolsep{2pt}\bea y&=&z + \alpha\sum_{m=1}^{M} p_{m} b_{m} \notag \\
&=&z + 4\alpha\sum_{m=1}^{M} \tilde{p}_{m} \tilde{b}_{m} +2\alpha\sum_{m=1}^{M} \tilde{p}_{m}.\eea}
Since $|z|\!<\!2$ and $\alpha\!>\!1$, it holds that $z\!+\!2\alpha\!>\!0$. If $\frac{1}{2}\sum\limits_{m=1}^M \!p_{m}\!=\!\sum\limits_{m=1}^M \tilde{p}_{m}$ is odd, we have
\bea  \label{A1} y \;\,\mathrm{mod} \;\,4\alpha=z+2\alpha; \eea
Otherwise, if $\sum\limits_{m=1}^M \tilde{p}_{m}$ is even, it also holds that
\bea \label{A2}  (y+2\alpha) \;\,\mathrm{mod} \;\,4\alpha=z+2\alpha. \eea
Combing (\ref{A1}) and (\ref{A2}), $z$ can be obtained as in (\ref{z1}).

\section*{Appendix B: A $4\!\times\!4$ example for applying the MZF detection}
Below we give a $4\!\times\!4$	real-valued MIMO example with 4-PAM modulation to illustrate the process of MZF detection, with assuming the channel, transmitted symbol vector and received signal vector as
\bea \vec{H}\!=\!\left[\!\begin{array}{cccc} -6&0&-1&5\\-3&-2&-1&1\\1&-5&-6&0\\1&-1&-3&-2\end{array} \!\right]\! \!, \, \vec{x}\!=\!\left[\!\begin{array}{rrrr} 1\\-1\\-1\\1\end{array} \!\right]\!\!,  \,  \vec{y}\!=\!\left[\!\begin{array}{rrrr} 3\\1\\15\\11\end{array} \!\right]\!\!, \notag
\eea
respectively. Then it can be shown that
\bea \vec{H}^{+}\!=\!\frac{1}{185}\left[\!\begin{array}{cccc} -5&-55&30&-40\\ 35&   -59&   -25& 58\\-30&    40&   -5&  -55\\  25&   -58&    35&   -59\end{array} \!\right]\!\!,\notag \eea
and the ZF estimate of $\vec{x}$ equals
\bea \tilde{\vec{x}}_{\mathrm{ZF}}=\vec{H}^{+}\vec{y}=\frac{1}{185}\left[\!\begin{array}{rrrr} -60\\309\\-730\\-107\end{array} \!\right]\!\!, \notag\eea
where only the third symbol is correctly detected.

Next we use the basic MZF detection with Algorithm 1. Setting $\tau\!=\!1$ and run SD for optimization (\ref{opti}) yields an optimal $\vec{Q}$ as
\bea \vec{Q}\!=\!\left[\!\begin{array}{cccc} -2&0&0&0\\ 0&   0&   2& 0\\0&    0&   -2&  0\\  2&   0&    0&   -2\end{array} \!\right]\!\!.\notag \eea
We first see that, the MZF shall reuse the ZF estimates for the first and third layers based on (\ref{badcond}). Then, we see that with $\vec{Q}$, the post-processing SNR (assuming the noise power equals 1) for the second bit-layer (which is identical to the fourth bit-layer) is increased from $1/\|\boldsymbol{\delta}_2\vec{H}^{\mathrm{+}}\|\!=\!185/47$ to $1/\|(\boldsymbol{\delta}_2+\vec{q}_2)\vec{H}^{\mathrm{+}}\|^2\!=\!185/27$. Next we compute estimates with the MZF for the these two layers.

For the second layer, according to (\ref{zkdef}) we have
\bea r_2=(\boldsymbol{\delta}_2+\vec{q}_2)\tilde{\vec{x}}_{\mathrm{ZF}}=\frac{1}{185}[ 0\;\; 1\;\; 2\;\;  0] \left[\!\begin{array}{rrrr} -60\\309\\-730\\-107\end{array} \!\right]\!=\frac{-1151}{185}, \notag \eea
and
\bea  z_2=(r_2 \;\,\mathrm{mod}\;\, 4) -2=\frac{-7}{38}. \notag \eea

Similarly, for the fourth layer we have
\bea r_4=(\boldsymbol{\delta}_4+\vec{q}_4)\tilde{\vec{x}}_{\mathrm{ZF}}=\frac{1}{185}[ 2\;\; 0 \;\;0\;\; -1] \left[\!\begin{array}{rrrr} -60\\309\\-730\\-107\end{array} \!\right]\!=\frac{-13}{185}, \notag \eea
and
\bea  z_4=(r_4 \;\,\mathrm{mod}\;\, 4) -2=\frac{357}{185}. \notag \eea
As can be seen, the MZF corrects both detections for the second and the fourth layers where the ZF fails.

\section*{Appendix C: Proof of Property 2}

In the case where each transmit antenna encodes an independent data stream, the optimal DMT  \cite{ZT03} is
\bea d_{\mathrm{ML}}(r)=N\left(1-\frac{r}{K}\right),\; 0\leq r\leq K, \eea
which can be achieved by joint maximum likelihood (ML) decoding, while the ZF (and also the LMMSE) receiver attains a DMT
\bea d_{\mathrm{ZF}}(r)=(N-K+1)\left(1-\frac{r}{K}\right),\; 0\leq r\leq K. \eea

Since $\tilde{\vec{H}}$ has i.i.d. Rayleigh entries, $\tilde{\vec{H}}$ is full column rank with probability 1, and so is $\vec{H}\in\mathbb{Z}^{2N\times 2K}$ which is the real-valued representation $\tilde{\vec{H}}$. To prove that the MZF detection can achieve the optimal DMT, we follows a similar proof in \cite{ZG14} for IF receiver, which builds on a result in \cite{TK072} that showed that, uncoded signaling coupled with LR can achieve the full diversity and with a multiplexing gain of zero.

We define a lattices $\Lambda$ and its dual $\Lambda^{\ast}$ generated by $2N\!\times\!2K$ matrix $\vec{H}$ and $2K\!\times\!2N$ matrix $\vec{H}^{+}$ as \cite{LS90}
{\setlength\arraycolsep{2pt}  \bea  \label{lattice1} \Lambda&=&\left\{\vec{H}\vec{p},\, \vec{p}\in\mathbb{Z}^{2K} \right\}, \\
 \Lambda^{\ast}&=&\left\{(\vec{H}\rmt)^{+}\vec{p},\, \vec{p}\in\mathbb{Z}^{2K} \right\},\eea}
\hspace{-1.4mm}respectively.  From \cite{AH06, LS90, L95}, the $i$-th successive minimum $\epsilon_{i}(\Lambda)$ of a lattice $\Lambda$ is defined as the smallest length $r$ (with respect to the Euclidean norm) such that, there are $i$ vectors in $\Lambda$ of length at most $r$ that are linearly independent (with respect to $\mathbb{R}$). By this definition, it holds that \cite{ZG14}
{\setlength\arraycolsep{2pt} \bea  
 \label{eps1} \epsilon_{1}^2(\Lambda)&=&\min_{\vec{p}\in\mathbb{Z}^{2K}} \|\vec{p}\vec{H}\rmt\|^2,  \\
 \label{eps2}  \epsilon_{2K}^2(\Lambda^{\ast})&=&\min_{\substack{\vec{P}\in\mathbb{Z}^{2K\times 2K} \\ \text{rank}(\vec{P})=2K}} \max_{1\leq k\leq 2K}\|\vec{p}_k\vec{H}^{+}\|^2.
\eea}
\hspace{-1.4mm}Then, we state the  following Lemma 1, which follows the proof in \cite[Appendix C]{ZG14}.
\begin{lemma}
Let $\Lambda$ be the lattice generated by $\vec{H}$
lattice generated by $\vec{H}$ according to (\ref{lattice1}). Then, for a sufficiently large $s\!>\!0$, the successive minimum $\epsilon_{K}^2(\Lambda)$ of $\Lambda$ satisfies
{\setlength\arraycolsep{2pt}\bea  \label{minima1} \mathbb{P}(\epsilon_{2K}^2(\Lambda^{\ast})>s)&\leq&\mathbb{P}\left(\frac{K^2(2K+3)}{\epsilon_{1}^2(\Lambda)}>s\right)  \\
\label{minima2} &=&\mathbb{P}\left(\epsilon_{1}^2(\Lambda)<\tilde{s}\right) \\
\label{minima3} &\leq& -\gamma \tilde{s}^{2N}(\ln\tilde{s})^{1+N} ,\eea}
where $\epsilon_{1}^2(\Lambda)$ is the first successive minimum for $\Lambda$, and
\bea  \tilde{s}=K\sqrt{\frac{2K+3}{s}}. \eea
\end{lemma}
The inequality (\ref{minima1}) follows from \cite[Lemma 4]{ZG14}, and the equality (\ref{minima2}) holds by switching $\epsilon_{1}^2(\Lambda)$ to the other side, and the inequality (\ref{minima3}) is a result of \cite[Lemma 5]{ZG14} when $\tilde{s}\!>\!0$ and is sufficiently small.

With the post-processing SNR $\gamma_k$ in MZF detection in (\ref{psnrMZF}), we can show that for a given target rate $r\ln\text{SNR}$, the outage probability equals
{\setlength\arraycolsep{1pt} \bea \label{ineq1} P(R<r\ln\text{SNR})&=&P\left(\max_{\vec{Q}\in2\mathbb{Z}^{2K\!\times \!2K}}\sum_{k=1}^{2K} \frac{1}{2}\ln(1+\gamma_k)<r\ln\text{SNR}\right)\notag \\  
&\leq&P\left(\max_{\substack{\vec{Q}\in2\mathbb{Z}^{2K\!\times \!2K} \\\text{rank}(\vec{Q})=2K }}\sum_{k=1}^{2K} \frac{1}{2}\ln(1+\gamma_k)<r\ln\text{SNR}\right)\notag \\
&\leq&P\left(\max_{\substack{\vec{Q}\in2\mathbb{Z}^{2K\!\times \!2K} \\\text{rank}(\vec{Q})=2K }} \min_{1\leq k\leq 2K}\gamma_k<\text{SNR}^{r/K}\right)\notag \\
&=&P\left(\max_{\substack{\vec{Q}\in2\mathbb{Z}^{2K\!\times \!2K} \\\text{rank}(\vec{Q})=2K }} \min_{1\leq k\leq 2K}\frac{\tau^2 \text{SNR}}{\|(\tau \boldsymbol{\delta}_k + \vec{q}_k)\vec{H}^+\|^2}<\text{SNR}^{r/K}\right)\notag \\
&=&P\left(\min_{\substack{\vec{Q}\in2\mathbb{Z}^{2K\!\times \!2K} \\\text{rank}(\vec{Q})=2K }} \max_{1\leq k\leq 2K} \|(\tau \boldsymbol{\delta}_k + \vec{q}_k)\vec{H}^+\|^2>\tau^2\text{SNR}^{1-r/K}\right)\!.
 \eea}
Using the inequality
\bea  \label{ineq2} \|(\tau \boldsymbol{\delta}_k + \vec{q}_k)\vec{H}^+\|^2 \leq  \|\vec{q}_k\vec{H}^+\|^2+\tau^2\|\boldsymbol{\delta}_k \vec{H}^+\|^2,   \eea
and inserting (\ref{ineq2}) back into (\ref{ineq1}) yields
{\setlength\arraycolsep{1pt} \bea \label{ineq3} P(R<r\ln\text{SNR})
&\leq&P\left(\min_{\substack{\vec{Q}\in2\mathbb{Z}^{2K\!\times \!2K} \\\text{rank}(\vec{Q})=2K }} \max_{1\leq k\leq 2K} \| \vec{q}_k\vec{H}^+\|^2>\tau^2\left(\text{SNR}^{1-r/K}-\|\boldsymbol{\delta}_k \vec{H}^+\|^2\right)\!\right)\notag \\
&=&P\left(\min_{\substack{\vec{P}\in\mathbb{Z}^{2K\!\times \!2K} \\\text{rank}(\vec{P})=2K }} \max_{1\leq k\leq 2K} \| \vec{p}_k\vec{H}^+\|^2>\frac{\tau^2}{4}\left(\text{SNR}^{1-r/K}-\|\boldsymbol{\delta}_k \vec{H}^+\|^2\right)\!\right)\notag \\
&=&P\left(\epsilon_{2K}^2(\Lambda^{\ast})>\frac{\tau^2}{4}\left(\text{SNR}^{1-r/K}-\|\boldsymbol{\delta}_k \vec{H}^+\|^2\right)\right)\!, \eea}
\hspace{-1.4mm}where the successive minimum $\epsilon_{2K}^2(\Lambda^{\ast})$ is defined in (\ref{eps2}). Comparing (\ref{ineq3}) to \cite[Eq. (43)]{ZG14}, it can be seen that the outage probability with MZF detection could be inferior to that with the IF receiver. However, the slope of the outage probability in high SNR regimes remains the same. Using Lemma 1, it holds that
{\setlength\arraycolsep{1pt}\bea  \label{limDMT}  d_{\mathrm{MZF}}(r)&=&\lim_{\text{SNR}\to\infty}\!\! \frac{-\ln P(R<r\ln\text{SNR})}{\ln\text{SNR}} \notag \\ &=& \lim_{\text{SNR}\to\infty}\!\! \frac{-\ln P\left(\epsilon_{2K}^2(\Lambda^{\ast})>\frac{\tau^2}{4}\left(\text{SNR}^{1-r/K}-\|\boldsymbol{\delta}_k \vec{H}^+\|^2\right)\right)}{\ln\text{SNR}}\notag \\
&\geq& \lim_{\text{SNR}\to\infty}\!\! \frac{-\ln\left(-\gamma \hat{s}^{2N}(\ln\hat{s})^{1+N}\right)}{\ln\text{SNR}}, \eea}
\hspace{-1.4mm}where
{\setlength\arraycolsep{1pt}\bea \label{hats}\hat{s}&=&\frac{2K}{\tau}\sqrt{\frac{2K+3}{\text{SNR}^{1-r/K}-\|\boldsymbol{\delta}_k \vec{H}^+\|^2}} \notag \\
&\geq&\frac{2K}{\tau}\sqrt{\frac{2K+3}{\text{SNR}^{1-r/K}}} .\eea}
\hspace{-1.4mm}Inserting (\ref{hats}) into (\ref{limDMT}), it holds that
\bea \label{dmzf1} d_{\mathrm{MZF}}(r) \geq N\left(1-\frac{r}{K}\right). \eea
But as the MZF detection is inferior to the ML detection, so is the DMT. Therefor, the equality in (\ref{dmzf1}) holds.
							
\bibliographystyle{IEEEtran}

\end{document}